\documentclass[12pt]{article}
\usepackage{url}

\usepackage{graphicx}
\usepackage{amsmath, amsfonts, amssymb, rotating, setspace}
\usepackage[utf8]{inputenc}
\usepackage[margin=2.5cm]{geometry}
\usepackage{url} 
\usepackage{natbib}
\usepackage{hyperref}
\usepackage{tikz}

\usetikzlibrary{calc}
\usetikzlibrary{positioning}
\usetikzlibrary{arrows}

\begin{document}
 
\markboth%
{M.C. Sachs and others}
{Flexible surrogates in platform studies}

\title{Flexible evaluation of surrogacy in Bayesian adaptive platform studies}
\author{Michael C Sachs$^{1,2,\ast}$ \and Erin E Gabriel$^{1,2}$ \and Alessio Crippa$^2$ \and Michael J Daniels$^3$}
\date{%
    $^1$Section of Biostatistics, University of Copenhagen, Denmark\\%
    $^2$Department of Medical Epi and Biostatistics, Karolinska Institutet, Sweden\\
    $^3$Department of Statistics, University of Florida, USA\\
    $^*$michael.sachs@sund.ku.dk\\[2ex]%
    \today
}

\maketitle

\begin{abstract}
{Trial level surrogates are useful tools for improving the speed and cost effectiveness of trials, but surrogates that have not been properly evaluated can cause misleading results. The evaluation procedure is often contextual and depends on the type of trial setting. There have been many proposed methods for trial level surrogate evaluation, but none, to our knowledge, for the specific setting of Bayesian adaptive platform studies. As adaptive studies are becoming more popular, methods for surrogate evaluation using them are needed. These studies also offer a rich data resource for surrogate evaluation that would not normally be possible. However, they also offer a set of statistical issues including heterogeneity of the study population, treatments, implementation, and even potentially the quality of the surrogate. We propose the use of a hierarchical Bayesian semiparametric model for the evaluation of potential surrogates using nonparametric priors for the distribution of true effects based on Dirichlet process mixtures. The motivation for this approach is to flexibly model relationships between the treatment effect on the surrogate and the treatment effect on the outcome and also to identify potential clusters with differential surrogate value in a data-driven manner. In simulations, we find that our proposed method is superior to a simple, but fairly standard, hierarchical Bayesian method. We demonstrate how our method can be used in a simulated illustrative example (based on the ProBio trial), in which we are able to identify clusters where the surrogate is, and is not useful. We plan to apply our method to the ProBio trial, once it is completed.}
\\{Trial-level surrogate; Surrogate markers; Biomarkers; Multilevel model; Bayesian nonparametrics; Prostate cancer.}
\end{abstract}

\section{Introduction}
Confirmatory randomized clinical trials should be based on clinical endpoints such as death or cure. However, these outcomes can take a long time to observe increasing both the cost and time to approval of new, and potentially effective, treatment strategies. Trial level surrogates, termed trial level predictive surrogates in \citet{gabriel2016comparing}, that have been properly evaluated can be used to decrease the length and cost of trials by being used as replacement endpoints. Such endpoints are particularly useful in Phase II and IIb trials, where new treatments of strategies are being evaluated prior to final rigorous testing in a Phase III study. \citet{gabriel2016comparing} and \citet{gabriel2019optimizing} both used examples from multiple trials to evaluate the Zoster vaccine; the trials had differing protocols, were conducted in different countries, and were approximately independent that allowed a trial level surrogate analysis. However, this was a massive data collection effort that is likely not feasible in most situations. 

Bayesian adaptive randomized trials are becoming more common in studies of cancer treatment \citep{meyer2020evolution} and will likely become even more common in chronic illness studies. As Bayesian adaptive trials become more common, methods that can be used to evaluate candidate surrogates using data from them are needed. These trials offer a useful resource because they include diverse patient populations and several treatments and treatment strategies in one trial, rather than over a set of independent trials. In our motivating setting of adaptive platform studies, most of the treatments evaluated will already be approved and the evaluation of a potential surrogate will be based on subgroups defined by a predictive signature based strategy for treatment. Given the breadth of treatments and subgroups in a platform study, when a surrogate is determined to be valuable overall, it is more likely to be useful in general for a clinical outcome compared to a surrogate evaluated for a specific treatment or in a given subpopulation. 


We take a meta-analytic approach to the evaluation of surrogates, which has a long literature including both Bayesian and frequentist methods \citep{daniels1997meta, Gail00, burzykowski2006evaluation, Baker06, Korn05, Li2010, Dai12}. As trial level surrogates are most often contextual, i.e. applying to a particular treatment or treatment class and for a particular outcome, we will extend the definition of trial level surrogate to include a set of treatments and biomarker-based population subgroups. Thus, in addition to the developing methods for the evaluation of a trial level surrogate over all treatments, treatment strategies, and biomarker subgroups that a given platform study might include, we also aim to leverage the structure of the platform study to identify biomarker subgroups and/or treatments where a surrogate is most useful, and those where it possibly does not work or is less useful; if such variation exists in the surrogate usefulness. 

\citet{gabriel2016comparing, gabriel2019optimizing} developed flexible surrogate evaluation methods that could accommodate variation in surrogate quality, but this was not the focus of either paper. Recently, \citet{papanikos2020bayesian} considered hierarchical models for surrogate evaluation with a focus on variation in surrogate quality and propose a fully exchangeable model and a partially exchangeable model; the simple model we compare to in the simulations is most closely related to the fully exchangeable method proposed in \citet{papanikos2020bayesian}. In the setting of Bayesian adaptive platform trials, with multiple treatments and biomarker groups, it becomes much more likely such variation may exist in practice. To our knowledge, there have been no methods for surrogate evaluation developed for use in Bayesian adaptive, or any randomization adaptive studies, or that consider variations of the surrogate quality in such settings. 

Our motivating trial is the ongoing ProBio study, a biomarker driven platform trial in which metastatic prostate cancer patients are randomized to either the biomarker driven arm or the standard of care arm \citep{crippa2020probio, de2022clinical}. In the biomarker driven arm, patients are randomized to one of a subset of several possible treatments based on combinations of cancer cell mutations. The primary outcome of the study is progression-free survival. A potential surrogate endpoint that is being considered is circulating tumor DNA (ctDNA) \citep{vandekerkhove2019circulating}. To evaluate ctDNA as a trial level surrogate we model the association between the treatment effect on ctDNA and the treatment effect on the clinical outcome over a set of trials or sub-studies. In ProBio, all treatment arms use the same control group, the standard of care, and the treatments are partially nested within biomarker groups, referred to as biomarker signatures. Thus, one of the key statistical challenge is to flexibly model the association, while accounting for the partially overlapping hierarchical structure of the biomarker signatures, and the small group sample sizes.

In this paper we propose a hierarchical Bayesian semiparametric model for the evaluation of potential surrogates in a platform trial using nonparametric priors for the distribution of the true effects based on Dirichlet process mixtures. This model allows for both borrowing information across subgroups while maintaining flexibility and not imposing strong parametric assumptions. This flexible data-adaptive approach is capable of modelling nonlinear and complex structures among the joint distribution of treatment effects on the potential surrogate, treatment effects on the outcome, and group-level covariates. Based on this approach we evaluate potential surrogates using a cross-validation framework as previously proposed in  \citet{gabriel2016comparing} and  \citet{gabriel2019optimizing}. After evaluating the overall surrogate performance, we examine subgroups defined by both covariates and the clusters identified by the model. 

The paper is organized as follows. In Section 2, we describe our modeling framework. In Section 3 we outline our procedure for sampling from the posterior distributions of interest. In Section 4 we describe our measures of surrogate value and methods for estimating them. In Section 5, our method is evaluated and illustrated using simulated data including a simulation to match what is expected in the ProBio trial. We close with a discussion of the approach and avenues for future work and application in Section 6.

\section{Methods}

\subsection{Notation} \label{notation}
Let $\boldsymbol{X}_i = (S_i, \tilde{Y}_i, \delta_i, W_{i1}, \ldots, W_{iK}, B_{i1}, \ldots, B_{iM})$ denote the vector of observed data for subject $i$, for $i = 1, \ldots, n$. $S_i$ is the continuous, positive ctDNA value, $\tilde{Y}_i$ is the observed time which is the minimum of the time to the clinical outcome of interest $Y_i$ and the censoring time, $\delta_i$ is an indicator that the observed time is the time to the clinical outcome of interest. The covariates $W_{ik} \in \{0, 1\}$ indicate whether subject $i$ received treatment $k$, such that if all $W_{ik} = 0$ then it indicates that subject $i$ was randomized to standard of care and otherwise only one of the $W_{ik}$ can be equal to 1. The covariates $B_{im} \in \{0, 1\}$ are indicator variables of whether the subject $i$ is in biomarker group $m \in \{1, \dots, M\}$. Exactly one of the $B_{im}$ has to be equal to one, as the biomarker groups are mutually exclusive. Denote $\boldsymbol{W}_i = (W_{i1}, \ldots, W_{iK})$ and $\boldsymbol{B}_i = ( B_{i1}, \ldots, B_{iM})$. We also have treatment/biomarker level information $Z_{k}, k = 1, \ldots, KM$, where each $Z_k$ is a vector of $d$ covariates observed for each of the groups defined by biomarker and treatment. These covariates might include information such as indicators of mechanism of action and dose, among other characteristics.  


\subsection{Model Specification} \label{modsect}

We model the observed data as follows. At the first stage, we assume parametric models for the conditional distributions of the potential surrogate given the treatment and biomarker indicators such that, for $j = 1, \ldots, KM$, $S_j | \boldsymbol{W}_j, \boldsymbol{B}_j, \boldsymbol{\theta}_s$ are independent normally distributed with variance $\sigma^2_s$ and expected value,
\begin{align}
E[S_i | \boldsymbol{W}_{i}, \boldsymbol{B}_{i}, \boldsymbol{\theta}_s] = 
\sum_{m = 1}^M \beta^B_{m} \cdot B_{im} +  \sum_{k = 1}^K \sum_{m = 1}^M \beta^{(BW)}_{mk} \cdot B_{im} \cdot W_{ik}, \label{smean}
\end{align}
where $$\boldsymbol{\theta}_s = (\beta^B_1, \ldots, \beta^B_M, \beta^{(BW)}_{11}, \ldots, \beta^{(BW)}_{MK}).$$ This is a saturated mean model for the effects of treatment within biomarker-treatment groups on the potential surrogate endpoint $S_i$.  

 We also assume a parametric model for the conditional distributions of the clinical outcome given the treatment and biomarker indicators such that, for $j = 1, \ldots, KM$, $\log(Y_j) | \boldsymbol{W}_j, \boldsymbol{B}_j, \boldsymbol{\theta}_y$ are independent normally distributed with variance $\sigma_y^2$, and expected value, 
\begin{align}
E[\log(Y_i) | \boldsymbol{W}_{i}, \boldsymbol{B}_{i}, \boldsymbol{\theta}_y] = 
 \sum_{m = 1}^M \gamma^B_{m} \cdot B_{im} + \sum_{m = 1}^M\sum_{k = 1}^K\gamma^{(BW)}_{mk} \cdot B_{im} \cdot W_{ik}, \label{ymean}
\end{align}
where $\boldsymbol{\theta}_y = (\gamma^B_1, \ldots, \gamma^B_M, \gamma^{(BW)}_{11}, \ldots, \gamma^{(BW)}_{MK})$. This is again a saturated mean model (on the log scale) for the effects of treatment within biomarker-treatment groups on the clinical endpoint $Y_i$.

Our parameters of interest are the pairs of parameter vectors that quantify the contrasts comparing each treatment to the standard of care within each biomarker group which we denote as $(\boldsymbol\nu, \boldsymbol\mu)$ where $\boldsymbol{\nu} = (\nu_1 = \beta^{(BW)}_{11}, \nu_2 = \beta^{(BW)}_{12}, \ldots, \nu_{KM} = \beta^{(BW)}_{KM})$ and $\boldsymbol{\mu} = (\mu_1 = \gamma^{(BW)}_{11}, \mu_2 = \gamma^{(BW)}_{12}, \ldots, \mu_{KM} = \gamma^{(BW)}_{KM})$. The remaining parameters are the coefficients for the biomarkers terms (main effects of biomarkers) in the mean models \eqref{smean} and \eqref{ymean}. In what follows, we denote $\boldsymbol\theta_s = (\beta_0^*, \beta^B_1, \ldots, \beta^B_M, \boldsymbol\nu) = (\boldsymbol\eta, \boldsymbol\nu)$ and $\boldsymbol\theta_y = (\gamma^*_0, \gamma^B_1, \ldots, \gamma^B_M, \boldsymbol\mu) = (\boldsymbol\xi, \boldsymbol\mu)$. 

The likelihood is given as,
\[
p(\boldsymbol{X} | \boldsymbol\theta_s, \sigma_s, \boldsymbol\theta_y, \sigma_y) = \prod_{i = 1} ^n p(S_i | \boldsymbol{W}_i, \boldsymbol{B}_i, \boldsymbol\theta_s, \sigma_s) p(Y_i | \boldsymbol{W}_i, \boldsymbol{B}_i, \boldsymbol\theta_y, \sigma_y),
\]
where $\boldsymbol{X}$ is the observed data for all $n$ subjects as specified in Section \ref{notation}. The outcome $Y_i$ in our motivating example is the time to tumour progression or death. Due to the motivating trials being run in a country with a national health register, death will always be observable if it occurs during the trial period. Thus if one simply defines the primary outcome as time to observed progression or death, there will only be administrative censoring. 

In the second level of our model, we model the association between $\boldsymbol\nu$ and $\boldsymbol\mu$, the treatment effect on the true clinical outcome and the true treatment effect on the surrogate endpoint, conditional on group-level covariates. The standard (parametric) choice for the second level is a multivariate normal distribution, 
\[
(\nu_j, \mu_j)|\boldsymbol\omega, Z_j \sim_{iid} N\left(\left[\begin{array}{c}
     a_1 + b_1 Z_j  \\
     a_2 + b_2 Z_j
\end{array}\right], \Omega\right), \mbox{ where } \boldsymbol\omega = (a_1, a_2, b_1, b_2, \Omega).
\]
Instead, we adopt a Dirichlet process mixture (DPM) of multivariate normals for the joint distribution of $(\nu_j, \mu_j, Z_j)$, 
\[
(\nu_j, \mu_j, Z_j)|\omega_j \sim_{iid} G(\nu_j, \mu_j,  Z_j| \omega_j), j = 1, \ldots, KM,
\]
where $G$ is the kernel of a multivariate normal distribution of dimension $d+2$ (recall $d$ is the dimension of $Z_j$) with mean $\phi_j$, variance $\Sigma_j$, and  
\[
\omega_j = (\phi_j, \Sigma_j) \sim F, F \sim \mbox{DP}(\alpha, G_0),
\]
where $\alpha$ is the concentration parameter and the base measure, $G_0$ is a conjugate normal-inverse-Wishart (NIW) distribution. This specification allows for a flexible relationship of the treatment-level covariates $Z$ with the treatment effects on the surrogate and on the outcome \citep{muller1996bayesian}.  

For the hyper-parameters of the NIW distribution, we use the mean of the initial estimates weighted by their sample sizes for the location parameter, a diagonal matrix of the inverse variances of the initial estimates weighted by their sample sizes for the inverse scale matrix, $d+2$ for the degrees of freedom and $1/KM$ for the inverse scale parameter. We assume a Gamma$(a,b)$ prior on $\alpha$. 

We put noninformative priors on the main effects (nuisance parameters) of the biomarkers on the outcomes and surrogates
\[
\boldsymbol\eta, \boldsymbol\xi \sim_{iid} N(0, \tau),
\]
for a fixed parameter $\tau$ that is large. 
The variance parameters $\sigma_y \sim \mbox{Gamma}$ and $\sigma_s \sim \mbox{Gamma}$ with fixed shape and rate parameters. 



\section{Estimation, approximations, and derivations} \label{model}

Let $\boldsymbol\theta = (\boldsymbol\theta_s, \boldsymbol\theta_y) = (\boldsymbol\nu, \boldsymbol\mu, \boldsymbol\xi, \boldsymbol\eta, \sigma_y, \sigma_s)$ denote the first level parameters, and $\boldsymbol\omega = (\omega_1, \omega_2, \ldots)$ denote the parameters at the second level. We need to sample from the posterior distribution, $p(\boldsymbol\theta, \boldsymbol\omega | \boldsymbol{x}, z)$, from which we can then compute the posterior distribution,  $p(\boldsymbol\nu, \boldsymbol\mu | \boldsymbol{x}, z)$ and the posterior conditional distribution, $p(\boldsymbol\mu | \boldsymbol\nu, \boldsymbol{x}, z)$.

To sample from the posterior, we adapt Algorithm 8 of \citet{neal2000markov} to the hierarchical model specified in Section \ref{modsect}. Initial values are obtained by fitting regression models \eqref{smean} and \eqref{ymean} and setting the number of clusters to a value between 1 and $KM$ (e.g., determined using K-means clustering) and randomly sampling from the base measure. 

Let $\theta^{(u)}$ denote the current state of the sampling procedure. We implement the following steps:
\begin{enumerate}
    \item For $i = 1, \ldots, KM$: Let $k^-$ denote the number of distinct $\omega_j$ for $j \neq i$ and $h = k^- + 1$. Label the $\omega_j$ with cluster labels $c_j$ taking values in $\{1, \ldots, k^-\}$. Draw a new value $\omega_h$ independently from $G_0$. If $c_i \neq c_j$ for all $j \neq i$, let $c_i$ have the label $k^- + 1$. Draw a new value for $c_i$ from $\{1, \ldots h\}$ using these probabilities: 
    \begin{align*}
        \delta\frac{KM_{-i,c}}{KM - 1 + \alpha} G(\nu_i, \mu_i, Z_i| \omega_c),  \mbox{for } c < k^- + 1 \\
        \delta\frac{\alpha}{KM - 1 + \alpha} G(\nu_i, \mu_i, Z_i| \omega_c), \mbox{for } c = k^- + 1,
    \end{align*}
    where $\delta$ is the normalizing constant and $KM_{-i,c}$ is the number of $c_j = c$, for $j \neq i$. At the end of this step, we have an updated number of clusters $k$ with labels $\{c_1, \ldots, c_{KM}\}$. 
    \item For $i = 1, \ldots, k$, update the parameters $\{\omega_{1}, \ldots, \omega_{k}\}$ by Gibbs sampling,
    \[
    \omega_i | \omega_{-i}, \nu_i, \mu_i, Z_i \sim \sum_{i\neq j} q_{i,j} \delta(\omega_j) + r_i H_i, 
    \]
    where $H_i$ is the multivariate normal posterior of $\omega$ given the single observation $\nu_i, \mu_i, Z_i$, the ``likelihood'' $G$, and the conjugate prior $G_0$, and 
    \begin{align*}
        q_{i,j} & =  b G(\nu_i, \mu_i, Z_i | \omega_j) \\
        r_i & = b \alpha \int G(\nu_i, \mu_i, Z_i | \omega) \, dG_0(\omega)
    \end{align*}
    where $b$ ensures that $\sum_{i\neq j} q_{i,j} + r_i = 1$.
    \item Given the set of updated $\{\omega_{1}, \ldots, \omega_{k}\}$ from step 2, we update $\alpha$ as follows: (1) draw a value $z$ from the Beta$(\alpha + 1, n)$ distribution; (2) define $\pi_1 = a + k + 1, \pi_2 = n(b - \log(z))$ and $\pi = \pi_1/(\pi_1 + \pi_2)$; (3) Draw a new value $\alpha$ from Gamma$(a + k, b - \log(z))$ with probability $\pi$, or from Gamma$(a + k - 1, b - \log(z))$ with probability $1 - \pi$. 
    \item The values of $\boldsymbol\omega$ obtained in step 2 can be viewed as priors on different subsets of $(\boldsymbol\nu, \boldsymbol\mu, Z)$, one for each of the number of clusters $k$. However, to update the first stage models we need to condition on $Z$, assuming that $(Y, S) \perp Z | (\boldsymbol\nu, \boldsymbol\mu)$. Hence we derive the values of $(\nu_i, \mu_i) | Z_i$ for each $i = 1, \ldots, k$ from the multivariate normal distribution of $(\nu_i, \mu_i, Z_i)$.
    \item In the first stage models, if $\nu_j, \mu_j$ belongs to cluster $c_j$, the prior for those parameters is normal with parameters defined by the conditional distribution of $\nu_{c_j}, \mu_{c_j} | Z_{c_j}$. Combining these with the noninformative priors on $\boldsymbol\eta, \boldsymbol\xi$, and the Gamma priors for $\sigma_s, \sigma_y$, we then sample from the parameters in the regression models \eqref{smean} and \eqref{ymean} $(\boldsymbol\theta_s, \boldsymbol\theta_y)$ using Gibbs sampling.  
    \item Repeat steps 1-5. 
\end{enumerate}


\section{Surrogate quality evaluation} \label{sect:quality}

Our measure of surrogacy is the degree to which the treatment effects on the surrogate $\boldsymbol\nu$ are predictive of the treatment effects on the outcome $\boldsymbol\mu$. In particular, we are interested in computing $p(\tilde{\mu}_{KM+1} | \nu_{KM+1}, \boldsymbol{x}, z_{KM+1})$, where $\tilde{\mu}_{KM+1}$ denotes the unknown (i.e., outcome data is unavailable for the index ${KM+1}$) treatment effect on the outcome and $\nu_{KM+1}$ denotes the corresponding treatment effect on the surrogate for which we have data. We can express the desired conditional posterior predictive distribution as 
\begin{align} 
& p(\tilde{\mu}_{{KM+1}} | \nu_{KM+1}, \boldsymbol{x}, s_{KM+1}, \boldsymbol{z}, z_{KM+1}) = \nonumber  \\ 
& \int p(\tilde{\mu}_{KM+1} | \boldsymbol\theta, \boldsymbol\omega, \nu_{KM+1}, \boldsymbol{z}, z_{KM+1}) p(\boldsymbol\theta^{-\mu_{KM+1}}, \boldsymbol\omega | \boldsymbol{x}, s_{KM+1}, \boldsymbol{z}, z_{KM+1}) \, d\boldsymbol\theta^{-\mu_{KM+1}} \, d\boldsymbol\omega,  \label{predeq}
\end{align}
where $\boldsymbol{x}$ only includes data from groups $1, \ldots, KM$, $\boldsymbol{z}$ is the vector of group level covariates for groups $1, \ldots, KM$, $s_{KM+1}$ denotes the individual level data on the surrogate for group $KM+1$, $z_{KM+1}$ denotes the group-level data for group $KM+1$, and $\boldsymbol\theta^{-\mu_{KM+1}}$ denotes the vector of parameters $\boldsymbol\theta$ with $\mu_{KM+1}$ removed. 

The distribution of absolute differences between $\tilde{\mu}_{KM+1}$ and $\mu_{KM+1}$, denoted by $D$, quantifies the surrogate quality \citep{gabriel2016comparing}. If we knew the true value of $\mu_{KM+1}$, we could measure the value of our potential surrogate by comparing samples from it to the samples from the distribution of $\tilde{\mu}_{KM+1}$. Since we do not know the true value of $\mu_{KM+1}$, we can compute $D$ from the conditional posterior including all $KM$ groups, $(\boldsymbol\mu) | (\boldsymbol\nu, \boldsymbol{x}, \boldsymbol{z})$, and compare to the posterior conditional distribution \eqref{predeq} from a leave-one-out procedure. By iterating through $1, \ldots, KM$, we can get an estimate of how well the model predicts unseen treatment effects. 

More formally, we can estimate $D$, our measure of surrogate quality, by the following leave-one-out procedure after obtaining samples from the posterior of $(\boldsymbol\mu) | (\boldsymbol\nu, \boldsymbol{x}, \boldsymbol{z})$ using all of the available data.

For each $j \in \{1, \ldots, KM\}:$
\begin{enumerate}
    \item Set all $Y_{ij}$ in the individual-level data to missing. 
    \item Fit the model using the algorithm described in Section \ref{model}. 
    \item Generate samples from $\tilde{\mu}_j$, the leave-one-out posterior conditional distribution of $\mu_j$, by sampling from the distribution of $\mu_j | \nu_j, Z_j$ at step 4 of the algorithm given by equation \eqref{predeq}. 
    \item Compute $\hat{d}_j = |\tilde{\mu}_j - \mu_j|$ using a sample for $\mu_j$ based on the full data model posterior predictive. 
\end{enumerate}

For each posterior sample of parameters, $\hat{D} = \frac{1}{KM}\sum_j \hat{d}_j$ is an estimate of the distribution of the absolute prediction error for an unseen trial group. This is a measure of how well our model would predict unseen treatment effects on the outcome given information on the treatment effects on the potential surrogate and the group-level covariates. 

In order to assess the value of the potential surrogate, we compare $\hat{D}$ to how well we can predict the treatment effects on the outcome given only the treatment level covariates but not the treatment effect on the surrogate. We define $\hat{d}^{0}_j$ and $\hat{D}^{0}$ as the absolute prediction error based on a simplification of our model in Section \ref{modsect} so that the predictions of $\boldsymbol\mu$ are based on a model ignoring $\boldsymbol\nu$ and the models for the potential surrogate $S$. In particular, in our DPM, we have $(\mu_j, Z_j)|\omega_j \sim_{iid} G(\mu_j, Z_j | \omega_j)$ for $j = 1, \ldots, KM$. We refer to this as \emph{the null model}. Under this model, we can use a similar leave-one-out procedure to obtain samples from the distribution of $\overline{\mu}_j \sim p(\overline{\mu}_j | x, z_j)$. Comparing $\overline{\mu}_j$ to $\mu_j$ via the absolute prediction error $\hat{D}^0$ informs us how well the null model predicts unobserved treatment effects on the outcome. This null model incorporates information on $Z$ and the distribution of treatment effects on the outcome in a flexible manner, but ignores the distribution of treatment effects on the surrogate and its association with the treatment effects on the outcome. Comparing the distributions of $\hat{D}$ and $\hat{D}^0$ informs us about the value of the potential surrogate for predicting treatment effects relative to our best possible predictions in the absence of the potential surrogate information. The comparison can be based on summaries of each of the distributions, e.g., the mean or median, or we can estimate the posterior probability that the model incorporating the potential surrogate is better than the model without, $P(\hat{D} > \hat{D}^0 | \boldsymbol{x}, \boldsymbol{z})$.

In addition to assessing overall surrogate quality, it is of interest to determine whether there are subgroups based on the treatments, the biomarkers, or combinations thereof which have differential surrogate quality. To do this we can examine the distribution of $\hat{D}$ conditional on the treatments, biomarkers, and/or treatment-level covariates. The cluster labels of the DPM identify subgroups than can be interrogated for differential surrogate quality. We use the method of \citet{dahl2006model} to estimate the posterior cluster labels. The posterior samples of cluster labels are used to create a pairwise probability matrix for each group (the proportion of samples where groups are assigned to the same cluster), and then the cluster labels that minimize the sum of squared deviations of the probability matrix from the association assignment is used. 

\section{Simulation study}

The goal of the simulated examples is to assess the predictive performance of the DPM model compared to a fully parametric model which assumes a multivariate normal model at the second stage for $(\boldsymbol\nu, \boldsymbol\mu)$ with means linear in $Z$, the group-level covariates. The key distinction between our method and this comparator is the flexibility of the stage 2 model. 

\subsection{Data generation}

The structure of the simulated data is based on the simulations used in the power analysis of the ProBio trial\citep{crippa2020probio}. We specify a multinomial distribution for the 16 biomarkers, which represent the 16 combinations of 4 binary genetic markers as given in \citet{crippa2020probio}, which reports the main simulation study to assess the operating characteristics of the ProBio trial. For each of these 16 biomarker groups, there is randomization to 4 possible active treatments plus control. Hence for each simulation, there are 64 treatment effects (each active treatment compared to control for each biomarker group). We specified distributions of treatment effects on the potential surrogate conditional on a single group-level baseline covariate as standard normal. We generate the single group-level covariate $Z_j$ as normally distributed and an generate unobserved covariate $U_j$ also as standard normal, for $j = 1, \ldots, 64$. We also considered one scenario where the potential surrogate and group level covariates are distributed as skew-normal. 

We generate treatment effects on the clinical outcome $\mu_j$, given the treatment effects on the surrogate, $\nu_j$, according to the following models (the labels used in the tables are noted in parentheses): 

\begin{enumerate}
    \item Nonlinear (nonlinear): $\mu_j = -1 + f(\nu_j) + c_z |Z_j| + c_u U_j$, where $f$ is a linear spline basis with knots at $-1, 0, 1$ and coefficients chosen so that the trend is monotonically increasing.
    \item Linear (linear): $\mu_j = -1 + 1* \nu_j + c_z |Z_j| + c_u U_j$.
    \item Simple (simple): $\mu_j = -1 + 1* \nu_j + c_z Z_j + c_u U_j$.
    \item Null (null): $\mu_j = -1 +  c_z |Z_j| + c_u U$.
    \item Interaction (inter): $\mu_j = 1\{Z_j < 0\} * \nu_j + c_u U_j$, where $1\{\cdot\}$ is the indicator function. 
    \item Hidden interaction (interhide): $\mu_j = 1\{U_j < 0\} * \nu_j + c_z Z_j$. 
    \item Surrogate value for only 1 treatment (onetrt): $\mu_j = 1\{j\in R_a\} * \nu_j + c_z Z_j + c_u U_j$ where $R_a$ is the set of indices such that the active treatment is treatment A.
    \item Surrogate value for 2 treatments (twotrt): $\mu_j = 1\{j\in (R_a, R_b)\} * (\nu_j - \min_j(\nu_j)) + c_z Z_j + c_u U_j$ where $R_a, R_b$ are the sets of indices such that the active treatment is treatment A and B, respectively. Additionally, the mean of $Z_j$ is changed to vary based on the treatment group (with different means in each of the 4 groups). 
    \item Surrogate value for multiple biomarkers (manybiom): The biomarker groups are grouped into 3 categories $B^*_j \in \{1, 2, 3\}$ of size 20, 28, 16, respectively. Then $Z_j \sim N(B^*_j - 1, 0.5^2)$ and $\mu_j = 0.25 \cdot (B^*_j - 3) \cdot (\nu_j - \min_j(\nu_j)) + c_z Z_j + c_u U_j$. In this setting, the three biomarker groups have differential surrogate value none, moderate, and strong. 
\end{enumerate}

We consider the cases where $c_z \in \{0, .3\}$ and $c_u \in \{0, .3\}$ which correspond to null, moderate, and strong effects of the observed and unobserved group level variables on $\mu_j$. Details and summaries of these scenarios are available in the vignette of the R package that contains the code to reproduce the simulations. 

Once the treatment effects $(\nu_j, \mu_j)$ are generated, we generate individual-level data according to the simulation study described in \citet{crippa2020probio}. Briefly, the adaptive trial is run with randomization probabilities updated after each new 40 patients are enrolled, and stopping rules for benefit, futility and harm. Individual-level data are generated according to our stage 1 models with parameters determined by the treatment effects. For computational simplicity in the simulations, stopping rules for each biomarker-treatment combination are done in a simplified manner using a t-test. The trial continues for a fixed period of time, with randomization probabilities updated in the final periods to ensure that there are at least 2 individuals in each biomarker-treatment group. 

The sample sizes for the individual-level data for each of the 64 groups in each simulated trial were designed to match what is expected in the ProBio II trial. These do not differ much by setting nor by the values of $c_z$ and $c_u$ so they are only presented overall. In particular, the minimum sample size was 6, first quartile 10, median 13, third quartile 20 and maximum 115, on average over the replicates and settings. In the simulations reported in the main text, we generate only uncensored times to progression or death, as the main focus is on the evaluation of surrogate quality and the methods for handling censoring in the stage 1 model are standard. 

To demonstrate our method in a real data setting and to evaluate the performance with right censoring, we simulate a single example with censoring. In this single example we use the twotrt setting with $c_z = c_u = 0$, and generate censoring times as independent draws from a uniform distribution on $(20, 60]$ so that about 10\% of the individuals are censored. To accommodate censoring, we adapt our model by implementing the method to handle right censoring as described in \citep{qi2022bayesian}. 


\subsection{Analysis Implementation}

We implemented the computations for the DPM model as described in Section \ref{model} using the \texttt{dirichletprocess} R package \citep{dirichletprocess} and JAGS \citep{jags}. We also implemented the null model excluding the potential surrogate information, and the simple model which has a multivariate normal model at stage 2 also using JAGS. In these simulations, we used bivariate normal priors with mean 0, variance 2, and covariance $0.05$ for $\boldsymbol{\eta}, \boldsymbol{\xi}$. We used Gamma distribution priors with shape 1 and rate 1 for the variance parameters $\sigma^2_y$ and $\sigma^2_s$. 

Our measure of predictive performance is the absolute prediction error based on the leave-one-out estimates compared to the null model as described in Section \ref{sect:quality}. In addition, we report the posterior probability that each model has better predictive performance than the null model. The code for the simulation study and analysis methods are available in an R package called \texttt{dpsurrogate}, available on github (\texttt{sachsmc/dpsurrogate}) and in the supplementary materials. 

\subsection{Results}

Table \ref{dcomp} shows the means and standard deviations over the 100 simulation replicates of the  $\hat{D}$ (using the estimated treatment effects) and $D$ (using the true simulated treatment effects) using the DPM model and the simple hierarchical model. The DPM has superior predictive performance in all cases, and it appears that $\hat{D}$ is a good estimate of $D$. With larger values of the coefficients $c_z$ and $c_u$, the estimated values of $\hat{D}$ and $D$ tend to increase. The largest absolute performance gains for the DPM model over the simple model are in the observed interaction, null, and one treatment settings. 

Table \ref{pcomp} shows the mean and standard deviations over the simulation replicates of the posterior probability $P(\hat{D} < \hat{D}^0 | \boldsymbol{x}, \boldsymbol{z})$, that is, the probability that the surrogate has predictive value. With the exception of the simple settings, the DPM model has larger average probabilities of superiority than the simple model in all settings. In the null setting, where there is no surrogate value, the DPM still has substantially higher probability of superiority over the null than the simple model, though with the average probability of 0.21, this is not too high, but rather the probability of superiority for the simple model appears to be too low. The largest absolute probabilities are in the linear (but nonlinear in $Z_j$) and nonlinear settings, where the model is most easily able to detect differences from the null.

\begin{figure}
    \centering
    \includegraphics[width = .99\textwidth]{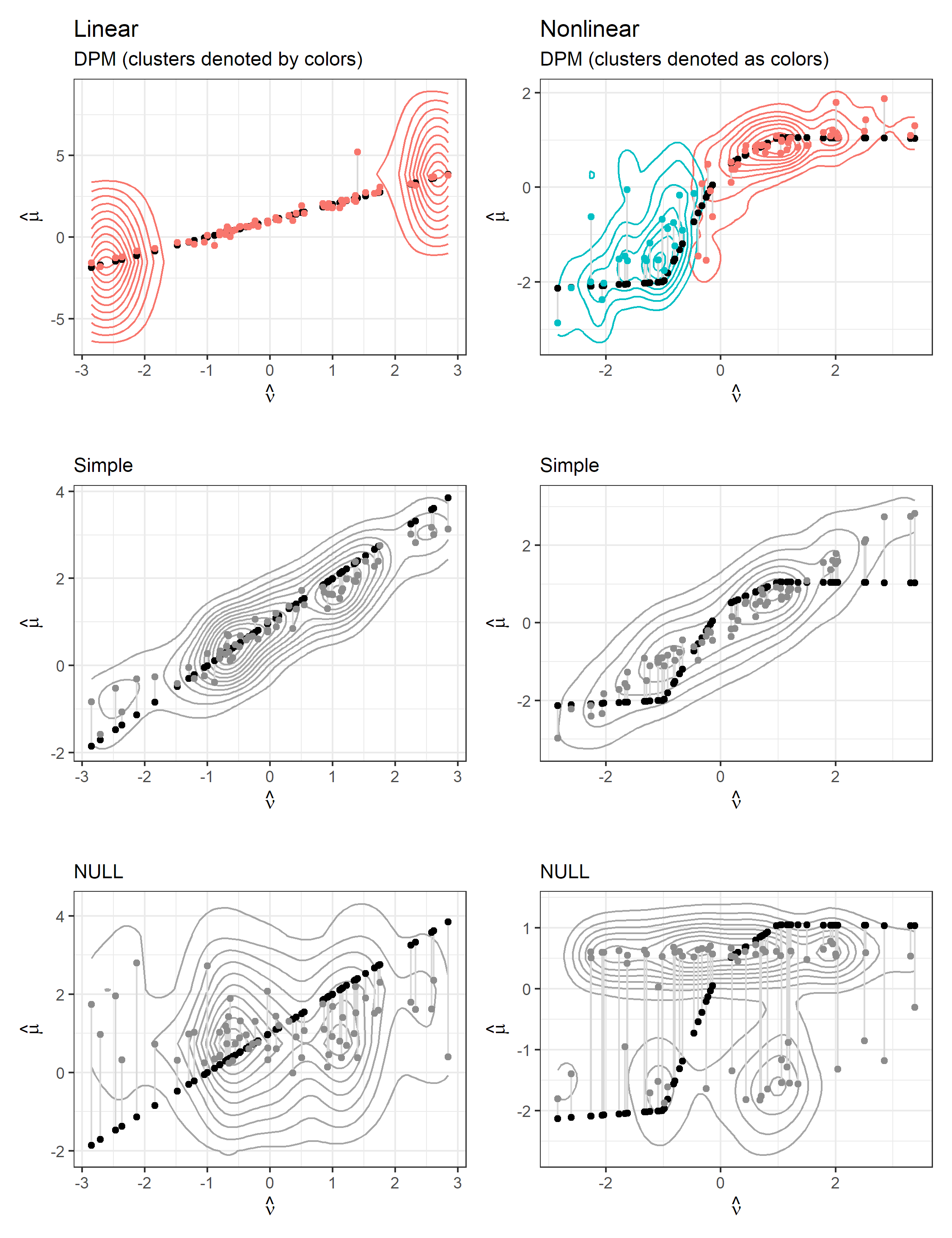}
    \caption{Medians of leave-one-out predictions versus true values, with contours for the posterior density.}
    \label{fig:med1}
\end{figure}

\begin{figure}
    \centering
    \includegraphics[width = .99\textwidth]{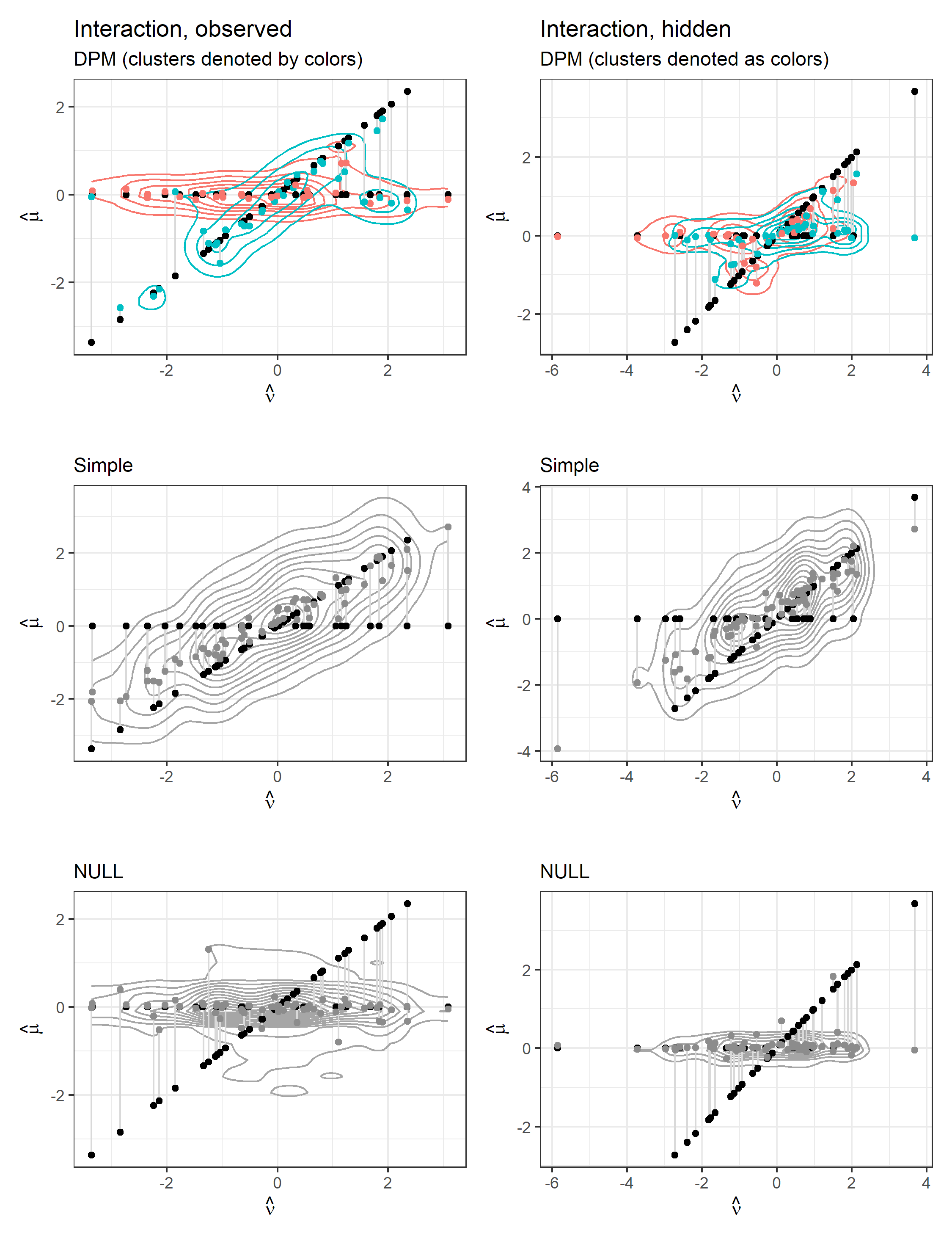}
    \caption{Medians of leave-one-out predictions versus true values, with contours for the posterior density.}
    \label{fig:med2}
\end{figure}

Figures \ref{fig:med1} and \ref{fig:med2} show a more detailed summary of the posterior distributions from an illustrative single replicate from the settings with $c_z = c_u = 0$. The DPM model is creating multiple clusters when appropriate, particularly in the nonlinear and observed interaction settings, to flexibly model complex associations. In the null setting, the DPM model tends to have more precise estimates of the treatment effect on the clinical outcome despite there being no association with the treatment effect on the potential surrogate (Figure S3). 

For the given sample size configuration, our proposed method has difficulty picking up the appropriate cluster in the one and two treatment settings and hence is not able to detect the improved value of the surrogate for the subgroups for which there is surrogate value (Figure S3), however, these are very difficult scenarios. When we consider variation over multiple biomarker groups, as in setting manybiom, we see that the cluster detection improves and we are able to detect the surrogate quality difference (Figure S6). Over the simulation replications in settings 7, 8, and 9, where there are clearly defined clusters, the mean and standard deviation of the proportion of clusters that are correctly identified for each leave-one-out group over the replicates is 0.76 (0.11) in the onetrt setting, 0.49 (0.06) in the twotrt setting, and 0.90 (0.05) in the manybiom setting. The DPM model is able to identify the correct cluster a large proportion of the time in the setting where there are distinct clusters. 

\begin{figure}
    \centering
    \includegraphics[width = .99\textwidth]{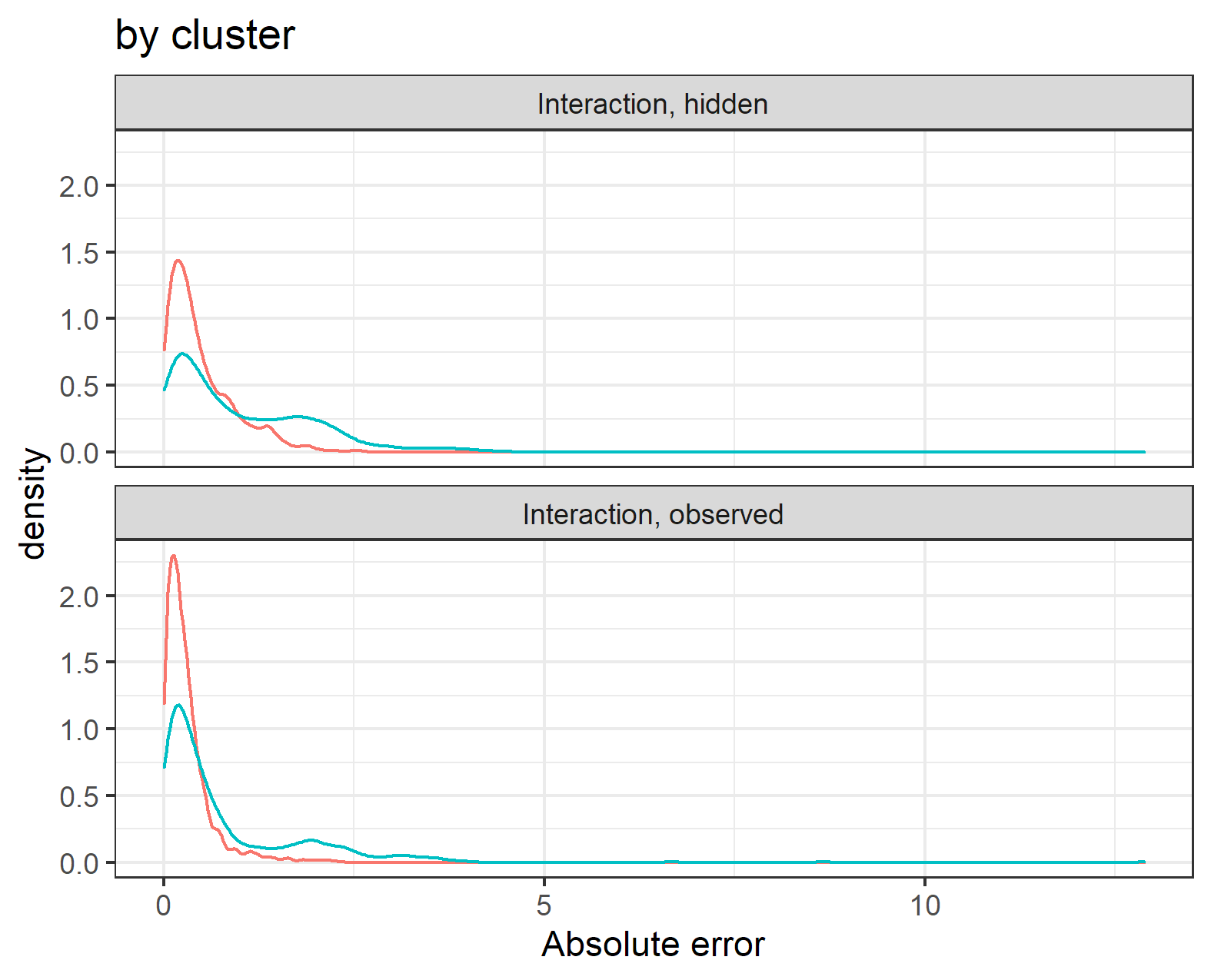}
    \caption{Posterior densities of leave one out absolute error with colors indicating the cluster assignments for the two interaction settings. In both cases, the red line indicates the density of the cluster with good surrogate value, and the blue line indicates the density of the cluster with poor surrogate value.}
    \label{fig:med3}
\end{figure}

Additionally, as can be seen in Figure \ref{fig:med3}, at least in some replicates, our method is not only able to identify the correct clustering based on surrogate quality that differs by observed subgroups (as in the inter settings), but also when this variation is based on unobserved latent variables (as in the interhide settings). Figure \ref{fig:med3} shows the posterior density of $\hat{D}$ for the two clusters, with the red line indicating the cluster with a higher quality surrogate relationship and the blue a lower quality surrogate relationship. The top panel is for the setting where the clusters are based on a latent variable and the lower panel for the setting where the clusters are based on a biomarker-treatment level observed covariate. In both settings, the DPM model is able to detect the differential surrogate value over the clusters. 

In summary, the simulation results show that we able to detect and estimate the quality of a surrogate that is useful over all treatments and biomarkers even when the surrogate effect to outcome effect relationship is complex. Our proposed semiparametric method better estimates the true surrogate quality, even when the true data generating mechanism closely matches a correctly specified parametric model. Additionally, we show our proposed method is able to detect variations in surrogate quality in most settings with adequate power. However, we see that to detect this variation on average we need a large enough subgroup with high surrogate quality, and a large difference in surrogate quality between subgroups (i.e., clearly defined clusters).

\subsection{Illustrative Example}
To illustrate how our proposed methods would be used in practice, from evaluation to the use of the evaluated surrogate in the next setting, we consider a single dataset generated under the two treatment (twotrt) setting, with $c_z = c_u = 0$, and with independent uniform censoring. Like all of the simulations scenarios, the data generation is based on the code used during the planning phase the Probio trial, and fits one of the scenarios that the trial team believed to be plausible. 

We assume that the clinical outcome is not yet observed in the group $j = 9$. Using our model, we would like to predict the unknown treatment effect on the clinical outcome in this group, based on the observed data in the 63 other treatment by biomarker groups in which both the candidate surrogate endpoint and the clinical outcome are observed. To do this, we fit our model using the available data, which does not include the clinical outcome for group $j = 9$, to obtain samples from the posterior of the treatment effect on the clinical outcome for that group. We compute estimates of the prediction error by running the leave-one-out procedure among the remaining 63 groups with complete data. 

\begin{figure}
    \centering
    \includegraphics[width = .99\textwidth]{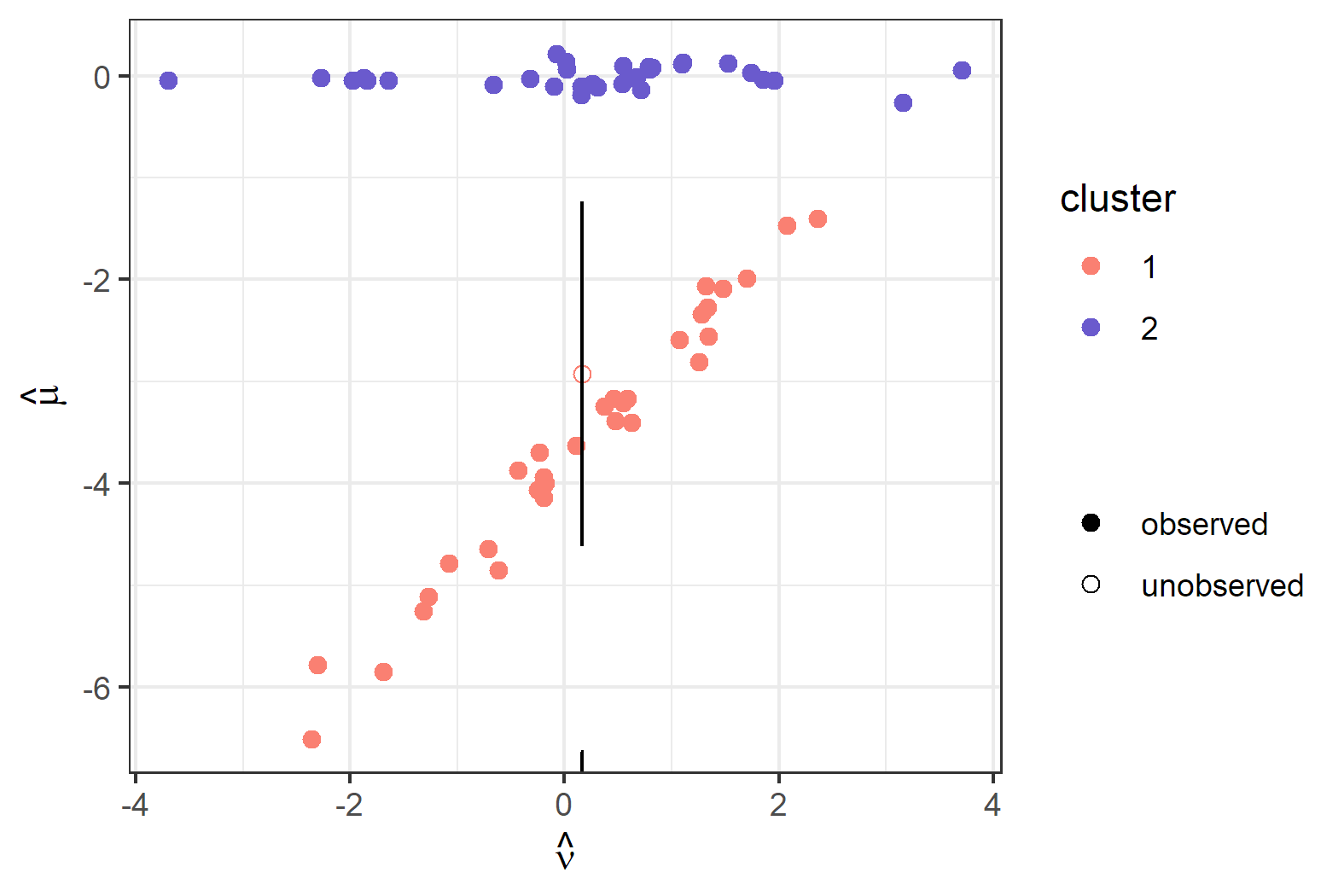}
    \caption{Illustrative example. The filled points represent the posterior medians of the treatment effects with fully observed data, while the unfilled point represents the posterior predicted median for the group where the clinical outcome is unobserved and the vertical line is a prediction interval obtained by adding and subtracting the median leave-one-out error in that cluster. }
    \label{figex}
\end{figure}

The results are shown in Figure \ref{figex}, which plots the posterior estimates of the treatment effects. The target is to predict the treatment effect on the clinical outcome for a new group where the clinical outcome in not measured, indicated by the open circle for the median prediction. This particular effect is correctly assigned to the cluster where there is high surrogate value, and over the iterations the cluster is correctly assigned 97.2\% of the time. 

The overall median prediction error of the DPM model is 2.02, but when looking by cluster, the median error is 2.29 in cluster 2, and 1.69 in cluster 1. In reference to the null model, which has an overall median prediction error of 2.23, the posterior probability of surrogate value $P(\hat{D} < \hat{D}^0 | \boldsymbol{x}, \boldsymbol{z})$ is 0.46 overall, 0.41 in cluster 2, and 0.53 in cluster 1. It is evident based on these estimates and in the figure that these clusters have differential surrogate value, and the model is able to correctly estimate cluster membership based on the surrogate and group-level covariates alone for the new group. 

For future use of the surrogate, one could fit the model with the new treatment effect on the surrogate to determine which cluster the new trial or group gets assigned to, then form posterior predictions of the treatment effect on the clinical outcome as we have done here. It would also be of interest to investigate how the treatment effect on the surrogate and the group level covariates determine cluster membership so that if there are observable determinants of the clustering, those could be used in future (rather than just rerunning the full model with the new treatment effect on the surrogate). In this example, the clusters are almost fully identified by two of the treatments. So, one could potentially rerun the model using only these two treatments and then use the resulting model moving forward.

\section{Discussion}
We propose a flexible and efficient model for assessing the value of a potential surrogate in the context of Bayesian adaptive randomized trials whose goal is to identify effective biomarker-treatment combinations. Our simulation study and example based on the ProBio trial demonstrates that this method is fit-for-purpose for evaluating the potential ctDNA surrogate in that context. The strengths of the method are the flexibility and data-adaptive nature of the priors in the hierarchical model. In addition the approach allows for baseline covariates to be conditioned on in a flexible manner. To our knowledge, this is a novel approach for surrogate evaluation. Although allowing for inclusion of covariates may make the surrogate quality look poorer, as these covariates may have some predictive ability, if these baseline measures are available this is likely a fairer estimate of the surrogate quality for a surrogate used in practice. 

Due to the clustering in our proposed method, we can assess surrogate quality variation over not only observed subgroups, but also data identified subgroups. Although simulations show that there are limits to how well this can work, particularly when the observed subgroups are very small or the surrogate quality varies only by a latent variable, we are able to detect variations in surrogate quality between large observed subgroups and in some settings with subgroups defined by latent variables. This is an important extension to previous work with the goal of identifying surrogate quality variation \citep{papanikos2020bayesian} or that allowed for such variation, without evaluating it, in a less flexible manner \citep{gabriel2016comparing, gabriel2019optimizing}. 

The shared control arm, adaptive nature of the trial, and stopping rules means that the treatment effect estimates based on the trial may be biased, especially for those that stopped due to superiority or futility \citep{emerson1990parameter}. Exploration of whether modifications to deal with such bias can be incorporated into our model is an area of future work. Our method specifies a parametric model in the first stage of the hierarchy and a nonparametric DPM at the second stage. Completely nonparametric hierarchical Dirichlet process mixtures have been developed and proven successful \citep{teh2006hierarchical}. Another avenue for future work would be to implement such mixtures to flexibly model the treatment effects at the first stage of the hierarchy as well. We have focused on a single continuous covariate. However, it will not be uncommon to have a mix of continuous and categorical $Z_j$. We can use the specification of \citet{shahbaba2009nonlinear} and assume independence among the categorical covariates within the DPM and/or in the case of many covariates, we can replace the DPM with an enriched DPM \citep{wade2011enriched}. Finally, although we demonstrate how independent censoring can be accounted for easily, investigation of dependent censoring may be useful for trial settings without registers and for outcomes other than all-cause death. 


\section*{Software}

Software in the form of an R package and complete documentation is available on
the corresponding author's GitHub at \url{https://github.com/sachsmc/dpsurrogate}.

\section*{Acknowledgements}
The authors report there are no competing interests to declare. MCS is partially supported by Swedish Research Council grant 2019-00227, EEG by Swedish Research Council grant 2017-01898, and MJD by National Institutes of Health grant R01 HL158963. 

\bibliographystyle{biorefs}
\bibliography{biblio}

\begin{table}[ht]
\centering
\begin{tabular}{lllll}
  \hline
setting &  DPM $\hat{D}$ & Simple $\hat{D}$ & DPM ${D}$ & Simple $D$ \\ 
  \hline
  inter-0.0-0.0 &  0.32 (0.04) & 0.80 (0.07) & 0.29 (0.04) & 0.79 (0.07) \\ 
  inter-0.0-0.3 &  0.45 (0.04) & 0.84 (0.07) & 0.44 (0.04) & 0.83 (0.07) \\ 
  interhide-0.0-0.0 &  0.50 (0.07) & 0.80 (0.07) & 0.48 (0.07) & 0.80 (0.07) \\ 
  interhide-0.3-0.0 & 0.49 (0.07) & 0.80 (0.08) & 0.46 (0.07) & 0.79 (0.08) \\ 
  linear-0.0-0.0 & 0.26 (0.02) & 0.40 (0.02) & 0.24 (0.02) & 0.39 (0.02) \\ 
  linear-0.3-0.0 & 0.30 (0.03) & 0.39 (0.02) & 0.29 (0.02) & 0.39 (0.02) \\ 
  linear-0.3-0.3 & 0.41 (0.04) & 0.48 (0.03) & 0.41 (0.03) & 0.47 (0.03) \\ 
  manybiom-0.0-0.0 & 0.44 (0.05) & 1.82 (0.15) & 0.42 (0.05) & 1.82 (0.14) \\ 
  nonlinear-0.0-0.0 & 0.43 (0.06) & 0.64 (0.05) & 0.40 (0.06) & 0.63 (0.05) \\ 
  nonlinear-0.3-0.0 & 0.45 (0.08) & 0.65 (0.07) & 0.43 (0.08) & 0.64 (0.07) \\ 
  nonlinear-0.3-0.3 & 0.54 (0.06) & 0.74 (0.06) & 0.53 (0.05) & 0.73 (0.06) \\ 
  nonlinearskew-0.3-0.0 & 0.41 (0.04) & 0.61 (0.05) & 0.40 (0.04) & 0.61 (0.05) \\
  null-0.0-0.0 & 0.51 (0.13) & 0.84 (0.04) & 0.49 (0.14) & 0.83 (0.04) \\ 
  onetrt-0.0-0.0 & 0.36 (0.07) & 0.87 (0.06) & 0.33 (0.07) & 0.87 (0.06) \\ 
  simple-0.0-0.0 & 0.25 (0.02) & 0.39 (0.02) & 0.24 (0.02) & 0.39 (0.02) \\ 
  simple-0.3-0.0 & 0.26 (0.02) & 0.40 (0.02) & 0.24 (0.02) & 0.39 (0.02) \\ 
  simple-0.3-0.3 & 0.38 (0.03) & 0.48 (0.03) & 0.37 (0.03) & 0.47 (0.03) \\ 
  twotrt-0.0-0.0 & 1.52 (0.90) & 2.09 (0.36) & 1.51 (0.92) & 2.09 (0.36) \\ 
   \hline
\end{tabular}
\caption{\label{dcomp} Mean and standard deviations of the median absolute prediction error based on leave-one-out CV. $\hat{D}$ denotes the estimate based on the full data DPM model, and $D$ in comparison to the true treatment effect.}
\end{table}

\begin{table}[ht]
\centering
\begin{tabular}{lllll}
  \hline
setting-$c_z$-$c_u$ & DPM $P(\hat{D} < \hat{D}^0)$ & Simple $P(\hat{D} < \hat{D}^0)$ & DPM $P({D} < {D}^0)$ & Simple $P({D} < {D}^0)$ \\ 
  \hline
inter-0.0-0.0 & 0.58 (0.04) & 0.43 (0.06) & 0.57 (0.05) & 0.42 (0.06) \\ 
  inter-0.0-0.3 & 0.57 (0.04) & 0.45 (0.05) & 0.56 (0.04) & 0.45 (0.05) \\ 
  interhide-0.0-0.0 & 0.52 (0.05) & 0.45 (0.06) & 0.50 (0.06) & 0.44 (0.07) \\ 
  interhide-0.3-0.0 & 0.52 (0.05) & 0.44 (0.06) & 0.51 (0.06) & 0.43 (0.06) \\ 
  linear-0.0-0.0 & 0.80 (0.03) & 0.84 (0.02) & 0.80 (0.03) & 0.84 (0.02) \\ 
  linear-0.3-0.0 & 0.81 (0.03) & 0.84 (0.02) & 0.81 (0.03) & 0.84 (0.02) \\ 
  linear-0.3-0.3 & 0.80 (0.02) & 0.81 (0.02) & 0.80 (0.02) & 0.81 (0.02) \\ 
  manybiom-0.0-0.0 & 0.62 (0.04) & 0.30 (0.02) & 0.62 (0.04) & 0.30 (0.02) \\ 
  nonlinear-0.0-0.0 & 0.72 (0.03) & 0.68 (0.05) & 0.72 (0.03) & 0.68 (0.05) \\ 
  nonlinear-0.3-0.0 & 0.72 (0.04) & 0.66 (0.05) & 0.72 (0.04) & 0.66 (0.05) \\ 
  nonlinear-0.3-0.3 & 0.69 (0.04) & 0.64 (0.04) & 0.70 (0.04) & 0.64 (0.05) \\ 
  nonlinearskew-0.3-0.0 & 0.72 (0.03) & 0.65 (0.05) & 0.72 (0.03) & 0.65 (0.05) \\ 
  null-0.0-0.0 & 0.21 (0.05) & 0.12 (0.01) & 0.14 (0.04) & 0.07 (0.01) \\ 
  onetrt-0.0-0.0 & 0.40 (0.04) & 0.26 (0.03) & 0.36 (0.05) & 0.24 (0.03) \\ 
  simple-0.0-0.0 & 0.80 (0.03) & 0.84 (0.02) & 0.81 (0.03) & 0.84 (0.02) \\ 
  simple-0.3-0.0 & 0.80 (0.04) & 0.84 (0.02) & 0.80 (0.04) & 0.84 (0.02) \\ 
  simple-0.3-0.3 & 0.80 (0.02) & 0.81 (0.02) & 0.80 (0.02) & 0.81 (0.02) \\ 
  twotrt-0.0-0.0 & 0.48 (0.04) & 0.39 (0.04) & 0.48 (0.04) & 0.38 (0.04) \\ 
   \hline
\end{tabular}
\caption{\label{pcomp} Mean and standard deviations of the posterior probability of superiority over the null model. $\hat{D}$ denotes the estimate based on the full data DPM model, and $D$ in comparison to the true treatment effect.}
\end{table}

\end{document}